# The Line Emission Terahertz Observatory (LETO): exploring the lifecycle of the ISM and the origins of water

Dimitra Rigopoulou[1], Peter Roelfsema[2,3], William Grainger [4], Chris Pearson[4], Boon-Kok Tan[1], Wouter Laauwen [2], Andrey Baryshev[3,5], Vincent Desmaris[6], Ismael Garcia-Bernete[7], Javier R. Goicoechea[8], Leslie Hunt[9], Georgios Magdis[10], Miguel Pereira-Santaella[8], Kamber Schwarz[11], Bryan Shaugnessy[4], Martina Wiedner[12] and the LETO Consortium

[1] University of Oxford, UK, [2] SRON, Netherlands, [3] University of Groningen, The Netherlands, [4] STFC RAL Space, [5] NOVA, The Netherlands, [6] GARD, Sweden, [7] CAB/CSIC, Spain, [8]IFF/CSIC, Spain, [9] INAF, Arcetri, Italy, [10] DTU, Denmark, [11] MPIA, Germany, [12] LIRA, France

## ABSTRACT

The Interstellar Medium (ISM) is the reservoir of baryonic matter from which stars and planetary systems are formed. It is also the repository of the material that is expelled at the end of the stellar evolutionary cycle feeding the baryonic matter reservoir. These evolutionary phase in the ISM together form a complex interplay driving planet and star formation and thus the evolution of our own Milky Way as well as galaxies at low and high redshifts. The Line Emission Terahertz Observatory (LETO) is designed to pin down the interrelation between stars and the ISM in much more detail than was possible ever before, by providing the capabilities to obtain extremely sensitive and high-resolution spectroscopic measurements in key Far-Infrared (FIR) molecular and atomic transitions.

LETO's design has been uniquely optimized to investigate the impact of the ISM on star formation on galactic and extragalactic scales, to study the processes that transform gas clouds into stars and planetary systems, and to trace the flow of water in the ISM, from galaxies to star-forming cores to planets and beyond. To achieve these goals, LETO will carry out deep velocity-resolved (3-Dimensional) wide-area surveys spectroscopic observations of key FIR lines in the ISM covering an area of approximately ~900 square degrees of the Galactic Plane. To complement our local view with earlier phases of cosmic evolution LETO will map a large sample of about 200 nearby galaxies in addition to surveys of Galaxies at Cosmic Noon. To shed light on the planet formation process, LETO will study the physical and chemical properties (especially gas mass) of previously known and newly discovered proto-planetary disks and stellar cores through e.g. pointed observations of the HD and $H_2O$ lines towards these sources.

LETO is a powerful Far-Infrared mission building on rich European heritage (Herschel/HIFI, GUSTO). To satisfy the requirements for sensitivity, resolving power and mapping speed, LETO will utilize a ~3.5m class mirror and several bands with sensitive state-of-the-art multi-pixel heterodyne arrays (4×4 goal). The bands together will cover a wide wavelength range (56–666μm) and with the heterodyne receivers and backends high resolving power ($R>10^6$) spectroscopy a set of key FIR atomic, ionic, and molecular lines can be studied in great detail. The wavelength bands are chosen such that they will include the two main HD transitions as well as a large suite of gas phase water lines. The mission is one of several selected for further study in the context of the ESA M8 call.

## 1. INTRODUCTION

Star formation is the fundamental driver of galaxy evolution, yet the physical processes that regulate the conversion of interstellar gas into stars remain poorly understood. The Interstellar Medium (ISM) is the reservoir of baryonic matter from which stars and planetary systems form. The ISM is a dynamic, multiphase environment shaped by gravity, turbulence, magnetic fields, stellar radiation, winds, supernovae, as well as the presence of active galactic nuclei (AGN). Understanding how these processes interact and regulate star formation (SF) —both in our Galaxy and across a range of external galaxies—is essential for a comprehensive picture of galaxy evolution. The advent of the Herschel Space Observatory [1] demonstrated the immense power of Far-Infrared (hereafter FIR) observations. The superb photometric capabilities of Herschel afforded panoramic views of large areas of the Galaxy by imaging the FIR dust emission (e.g. [2],[3],[4]). But these surveys provided a static view of the FIR emission without the velocity dimension that enables one to study the dynamics of the ISM.

To date, no large area velocity-resolved mapping of key far-infrared (FIR) lines of our Galaxy and external galaxies has been realised. LETO will address this challenge by delivering the first wide-field, velocity-resolved surveys of the key far-infrared (FIR) cooling lines—[C II] 158 μm, [O I] 63 μm and 145 μm, and [N II] 122 μm and 205 μm in our Galaxy — providing an unprecedented three-dimensional view of the cooling gas that regulates star formation and feedback.

In addition, velocity resolved observations of key FIR lines in nearby galaxies and of the redshifted OH 119 μm absorption feature at Cosmic Noon, will enable LETO to deliver the first statistical census of stellar and AGN feedback and directly measure quasar-driven molecular outflows during the peak epoch of galaxy assembly, establishing how feedback regulates star formation across cosmic time.

At the same time, LETO will directly probe the gas and volatile reservoirs from which planetary systems form. By conducting the first statistically significant survey of hydrogen deuteride (HD) in nearby protoplanetary disks and applying Doppler tomography at ultra-high spectral resolution ($R \approx 10^6$), the mission will measure disk gas masses and radial gas distributions, providing critical constraints on giant planet formation and migration. LETO will also trace the evolution of water and other volatiles from prestellar cores to disks and comets, revealing how these essential ingredients are transported and incorporated into planetary systems and, ultimately, potentially habitable worlds.

LETO uniquely combines a 3.5 m-class telescope, multi-pixel heterodyne arrays, wide instantaneous bandwidth, and ultra-high spectral resolution in the Terahertz regime. No previous or planned mission offers this capability. Building on Europe's leadership in far-infrared space astronomy, LETO will deliver a definitive legacy dataset:  three-dimensional atlas of the cooling ISM lines and the first direct, quantitative census of the gas and volatile reservoirs that govern planet formation. By resolving the ISM and planetary reservoirs in both space and velocity LETO will provide the missing dynamical link between galaxy evolution and the origin of planetary systems- establishing a unifying framework that connects the birth of stars to the emergence of worlds.

## 2. SCIENCE DRIVERS FOR THE LETO MISSION

### 2.1 From clouds to stars: the ISM in the Milky Way and External Galaxies

The scientific cornerstone of the LETO mission is the wide-area Galactic Plane Line Survey (GPLS). This program will deliver the first ever velocity-resolved reconstruction of the Milky Way's neutral and ionized ISM through observations of the [CII] 158 μm and [NII] 205 μm lines over 900 sq. deg. of the Galactic Plane. Figure 1 illustrates the scanning strategy that LETO will follow to obtain Nyquist-sampled maps of the 900 sq. deg in the Galactic Plane. The main goals of the survey are as follows:

First, GPLS will reveal the Galaxy's hidden molecular gas mass by quantifying the CO-dark molecular gas reservoir. High–spectral-resolution [C II] 158 μm mapping across the Galactic Plane, combined with existing H I and CO data, will isolate CO-dark $H_2$ gas velocity components, directly measuring their spatial distribution, their fraction relative to CO-bright gas, and dependence on Galactic environment. These results will establish the true baryonic budget of molecular gas in the Milky Way and provide a benchmark for interpreting molecular gas in low-metallicity galaxies and the early Universe.

Second, GPLS will measure stellar feedback on Galactic scales in action. By resolving the kinematics of expanding HII regions and bubbles in [C II] 158μm and [O I] 63μm, LETO will measure expansion velocities, kinetic energy injection rates, and the relative contributions of photoionization and stellar winds. This will clarify how massive stars regulate cloud lifetimes and drive turbulence in the surrounding ISM.

Third, GPLS will constrain turbulence injection and phase coupling. By measuring line profiles and velocity dispersions across multiple ISM phases, LETO will quantify Mach numbers and energy transfer between the diffuse and dense phases.

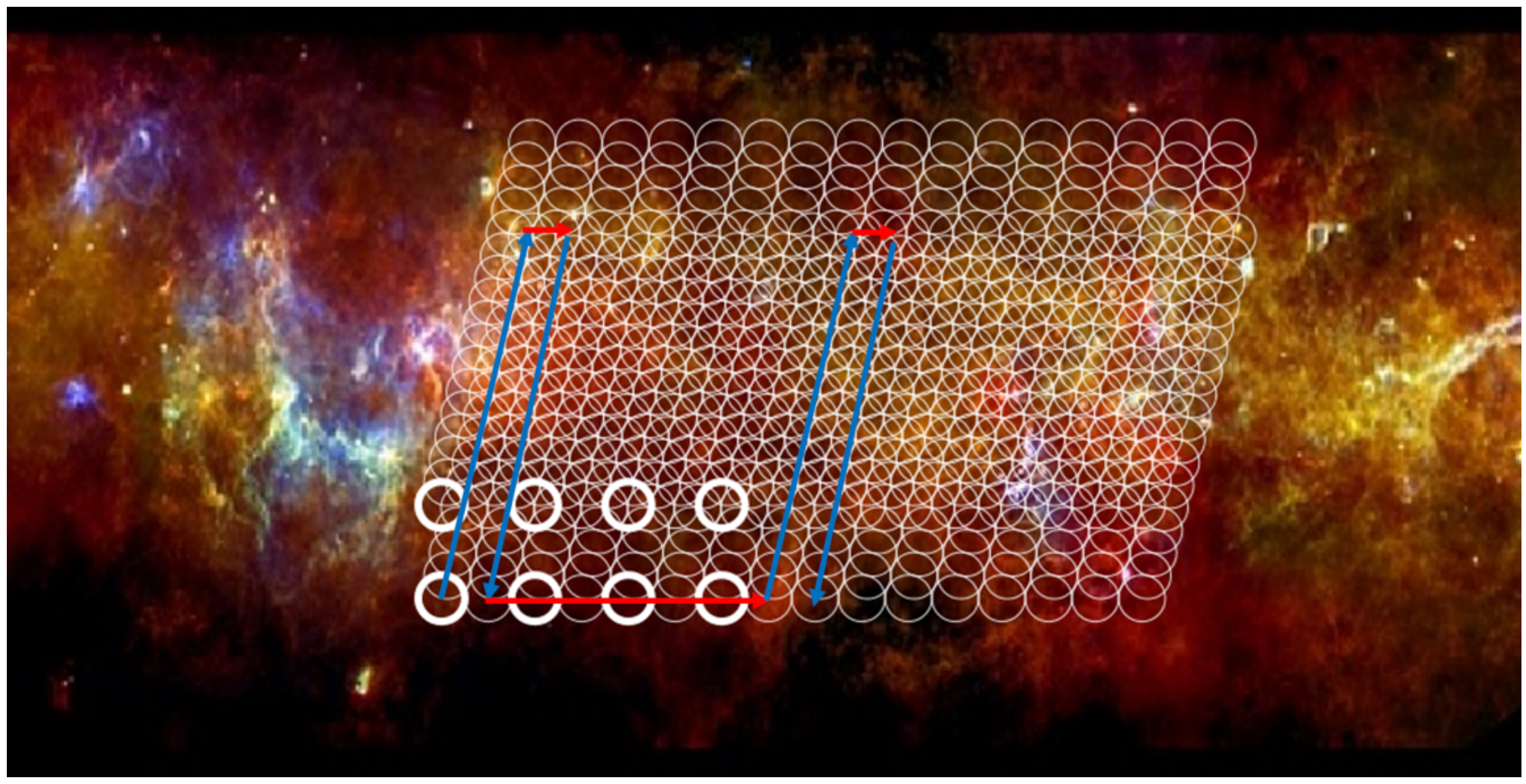

Figure 1: The LETO scanning strategy superimposed on a portion of Herschel's HiGAL continuum map (for demonstration purpose only not to scale).

These data will provide direct observational constraints on magnetohydrodynamic simulations of multiphase turbulence and establish how large-scale flows and feedback sustain supersonic motions.

Beyond the Milky Way, the Nearby Galaxies Line Survey (NGLS) will study approximately 200 galaxies spanning more than four orders of magnitude in star formation rate and over two orders of magnitude in metallicity. Velocity-resolved observations of [C II] 158μm, [O I] 63μm , and [N II] 205/122 μm lines will disentangle the ionized, atomic, and molecular contributions to FIR line emission, enabling a physically grounded interpretation of widely used star-formation tracers. The survey will establish the demographics of CO-dark molecular gas across different environments, recalibrate the CO-to-$H_2$ conversion factor, and determine the conditions responsible for the observed "[C II] deficit" in extreme systems such as ultraluminous infrared galaxies [5][6]. In addition, LETO will provide the first statistical census of multiphase stellar and AGN feedback in the local Universe by measuring gas kinematics associated with expanding super bubbles (45,46) galactic fountains, and large-scale outflows.

To connect local galaxy evolution with the distant Universe, LETO in addition will undertake the Cosmic Noon Galaxy Line Survey (CNGLS), targeting luminous AGN at redshifts around $z \approx 3$. Using velocity-resolved observations of the redshifted OH 119 μm transition, LETO will directly measure quasar-driven molecular outflows during the epoch of peak galaxy and black-hole growth. These observations will provide robust constraints on mass-loss rates and feedback efficiencies, establishing how AGN regulate star formation across cosmic time.

With its combination of wide-field mapping, high angular resolution, and spectral resolutions of a few km $s^{-1}$, LETO will deliver the first coherent, multiphase, three-dimensional view of the ISM from Galactic scales to the distant Universe. By quantifying hidden molecular gas reservoirs, measuring the energetics of stellar and AGN feedback, and establishing the physical basis of FIR diagnostics, LETO will transform our understanding of how galaxies form stars and evolve over cosmic time.

**2.2 From Stars to Planets: investigating planetary origins**

The second key objective of LETO is to directly measure the gas reservoir from which planets form. The bulk of disk mass in protoplanetary systems resides in molecular hydrogen ($H_2$), but $H_2$ is effectively invisible under typical disk conditions, forcing current studies to rely on indirect tracers such as dust continuum emission or CO, both of which introduce uncertainties approaching an order of magnitude [7][8]. LETO will conduct the first statistically significant survey of HD, the only direct tracer of the bulk hydrogen reservoir, through observations of the HD (1–0) and (2–1) transitions at 112 and 56 μm. These measurements will provide robust determinations of total disk gas masses, reducing the present uncertainties from factors of ~10 to approximately a factor of two. Such constraints are essential for understanding giant planet formation, pebble accretion, planetary migration, and the origin of the diverse exoplanet populations discovered over the last decade.

LETO's ultra-high spectral resolution ($\Delta\upsilon \approx 0.3$–$0.6$ km $s^{-1}$) will enable Doppler tomography of HD emission lines. Because protoplanetary disks are in Keplerian rotation, the velocity structure of spectral lines encodes radial information about the emitting gas. LETO will therefore reconstruct radial gas surface density profiles without spatially resolving the disks, providing the first direct measurements of how planet-forming material is distributed as a function of radius, stellar mass, and age. These observations will test whether sufficient gas exists at the locations and epochs required by current models of giant planet formation and migration. Complementary velocity-resolved observations of the [O I] 63 μm line will trace atomic disk winds, allow direct measurements of mass-loss rates and provide critical constraints on disk dispersal timescales and the available time for giant planet formation.

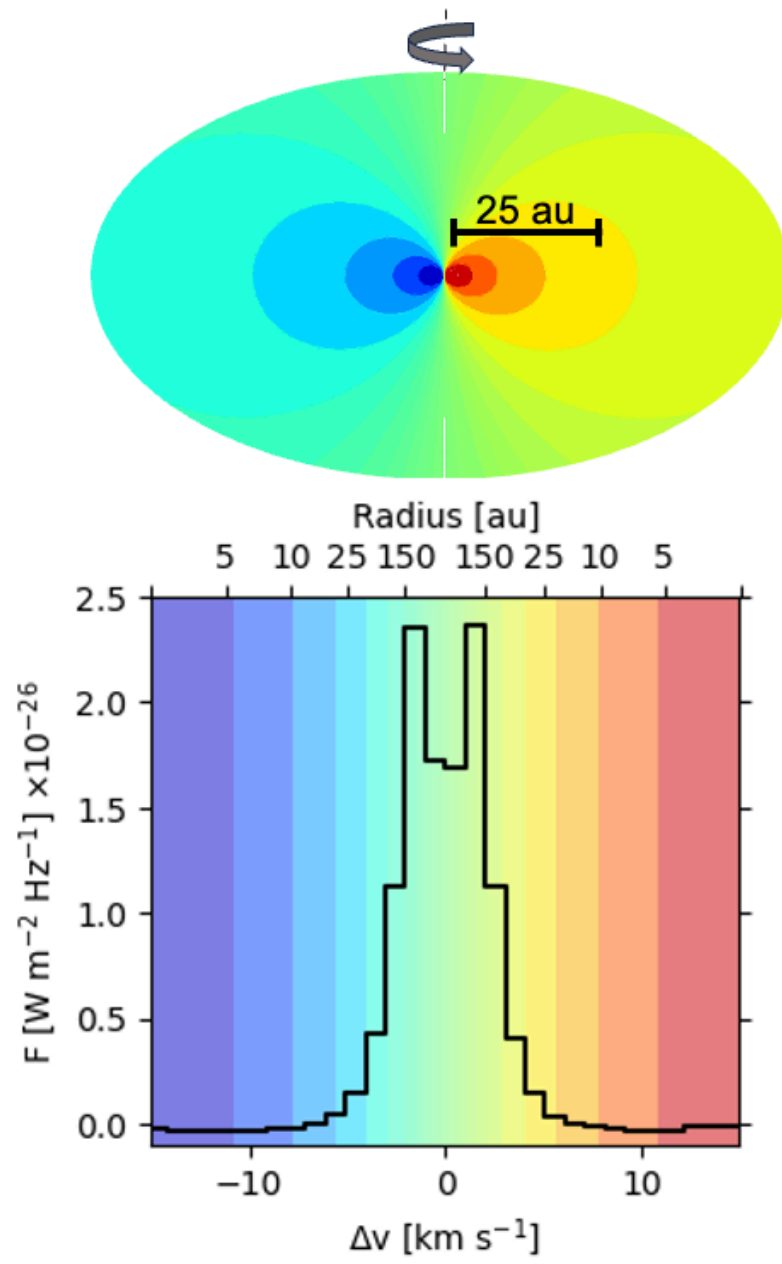

Figure 2: The Doppler tomography cartoon shows the expected spectrum for HD 1-0 is a minimum mass solar nebula disk around a solar mass star viewed at an inclination of 60 deg. and a distance of 140 pc

A second major theme is the origin and evolution of water and other volatile species that are central to planetary habitability. Water ice strongly enhances dust grain sticking efficiencies, promotes planetesimal growth, and influences the composition of emerging planets. However, while JWST has revealed substantial reservoirs of warm water in the inner regions of disks, [9][10] the cold outer-disk water reservoir remains largely unexplored because its most important transitions lie in the far-infrared. LETO will observe multiple water transitions simultaneously with HD, enabling measurements of water abundance and radial distribution throughout protoplanetary disks. Using Doppler tomography (Figure 2), LETO will distinguish emission arising from inner giant-planet-forming regions and outer disk reservoirs, constrain the location of snowlines, and directly probe the redistribution of icy material through pebble drift and radial transport. Observations of $H_2{}^{18}O$ and isotopic ratios such as HDO/$H_2O$ and ortho/para-$H_2O$ will further reveal the origin and thermal history of water, providing crucial links between interstellar chemistry and planetary composition [11]. LETO will also provide the first systematic constraints on the nitrogen budget in protoplanetary disks through observations of ammonia ($NH_3$), a potentially dominant nitrogen reservoir and a key ingredient in prebiotic chemistry. Simultaneous measurements of $NH_3$ and $H_2O$ will allow the mission to quantify the distribution of the two most important volatile carriers of nitrogen and oxygen, thereby improving our understanding of the elemental inventories available for planet formation.

Extending beyond disks, LETO will trace the complete "water trail" from the earliest stages of star formation to mature planetary systems. Deep observations of nearby prestellar cores will map the spatial distribution and kinematics of cold water vapor and ammonia, constraining the initial volatile inventory inherited by forming disks. In parallel, measurements of the deuterium-to-hydrogen ratio in cometary water will connect disk chemistry to Solar System bodies and test whether different comet populations preserve distinct records of volatile evolution. Together, these studies will establish the first observationally connected picture of how water is formed, processed, transported, and ultimately incorporated into planets.

## 3. THE LETO PAYLOAD

LETO's scientific objectives require sensitive spectroscopy across a wavelength range inaccessible from the ground due to atmospheric opacity. The mission targets velocity-resolved wide area observations of key ISM far-infrared cooling lines together with molecular tracers that constrain the gas and volatile reservoirs in planet-forming disks. Achieving these goals demands both high sensitivity and spectral resolving powers approaching $R \approx 10^6$, which can only be provided by cryogenic heterodyne instrumentation operating in space. The LETO payload therefore combines a 3.5-m class telescope with a suite of multi-pixel coherent SIS/HEB heterodyne receivers operating between 0.45 and 5.35 THz. Building strongly on the heritage of Herschel/HIFI [12], SOFIA/upGREAT [13], GUSTO [14], JUICE/SWI, [15] and ALMA, the payload extends proven single-pixel technologies into multi-pixel imaging arrays capable of large-scale spectroscopic mapping. The instrument architecture has been optimized to provide simultaneous access to key astrophysical tracers, including [C II], [N II], [O I], HD, $H_2O$, HDO, $NH_3$, and OH, while maintaining the spectral fidelity required for detailed kinematic studies. A comprehensive list of the lines covered by the LETO receivers can be found in Table 1, An artist impression of the LETO spacecraft is shown in Figure 3.

### 3.1 Payload Architecture

The LETO payload is composed of four principal elements: the LETO Telescope Assembly (LTA), the Cryo-Cooler Assembly (CCA), the cryogenic Focal Plane Unit (FPU), and the warm backend electronics comprising the Signal Processing Unit (SPU) and Instrument Control Unit (ICU). The observatory adopts a Herschel-like vertical telescope configuration in which the telescope optical axis is aligned with the focal-plane assembly. The receiver subsystem is positioned directly below the telescope, minimizing thermal distances between the cryogenic focal plane and the cooling system. Figure 5 presents a functional overview of the baseline LETO science instrument, with the cold optical bench units in the blue area on the top and in yellow the warm elements in the SVM on in the lower part of the diagram.

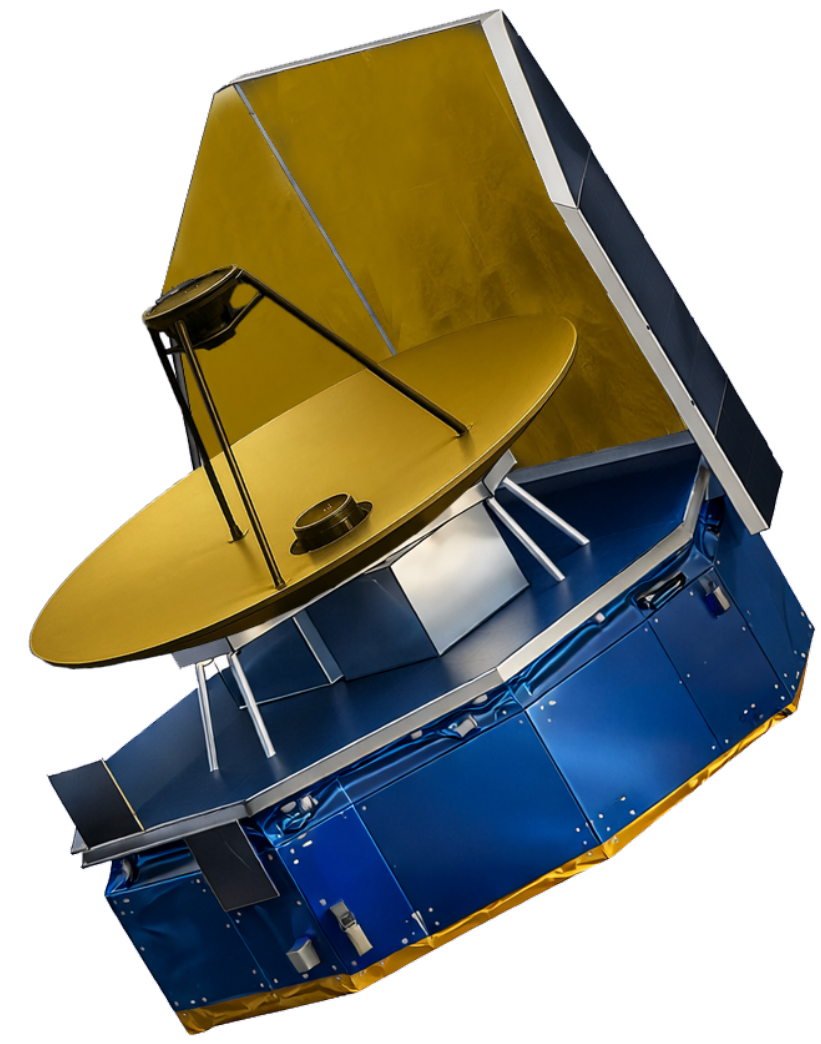

Figure 3 Artist impression of the LETO satellite. The payload – telescope and the cold instrument receivers – are shielded from the direct solar radiation. The 3.5meter telescope system is a direct copy of the Herschel design. The shield is clad with solar cells to provide power. The height of the shield allows up to 20° rotation around the Y-axis before the secondary is illumined by direct sunlight. The service module (SVM) including the instrument warm units is on the bottom.

A key advantage of the heterodyne approach is its relative insensitivity to thermal emission from the telescope itself. Consequently, LETO can operate with a warm telescope at approximately 100 K, avoiding the complexity associated with fully cryogenic telescope architectures while preserving the sensitivity required for the science goals. The observatory is designed for operation at the Sun–Earth L2 point, where a stable thermal environment and continuous sky accessibility maximize observing efficiency.

### 3.2 Telescope Assembly

The LETO telescope is based directly on the 3.5-m silicon-carbide telescope developed for the Herschel Space Observatory [16]. Reuse of this proven architecture significantly reduces technical risk while providing the collecting area required to detect faint ISM cooling lines and weak HD emission from protoplanetary disks.

The telescope employs a classical on-axis Cassegrain configuration with silicon-carbide primary and secondary mirrors mounted on a common low-expansion support structure. The optical design delivers angular resolutions ranging from approximately 37 arcsec at 500 GHz to 4.4 arcsec at 5 THz. The primary mirror is not actively temperature controlled but is expected to settle to a stable operating temperature of approximately 100 K through passive thermal equilibrium.

### 3.3 Receivers

The scientific payload is based on four independent heterodyne receiver bands optimized for specific science drivers. Together these bands provide continuous access to the spectral lines required for studies of the ISM, galaxy evolution, and planet formation. In the baseline concept the lowest frequency band 1 is equipped with Superconductor-Insulator-Superconductor (SIS) mixers while the higher frequency bands 2, 3 and 4 utilize Hot Electron Bolometer (HEB) mixers. The mixers all are single polarization sensitive, and to increase sensitivity each band has two sets of orthogonally polarized devices which are coaligned on the sky (see Figure 4).

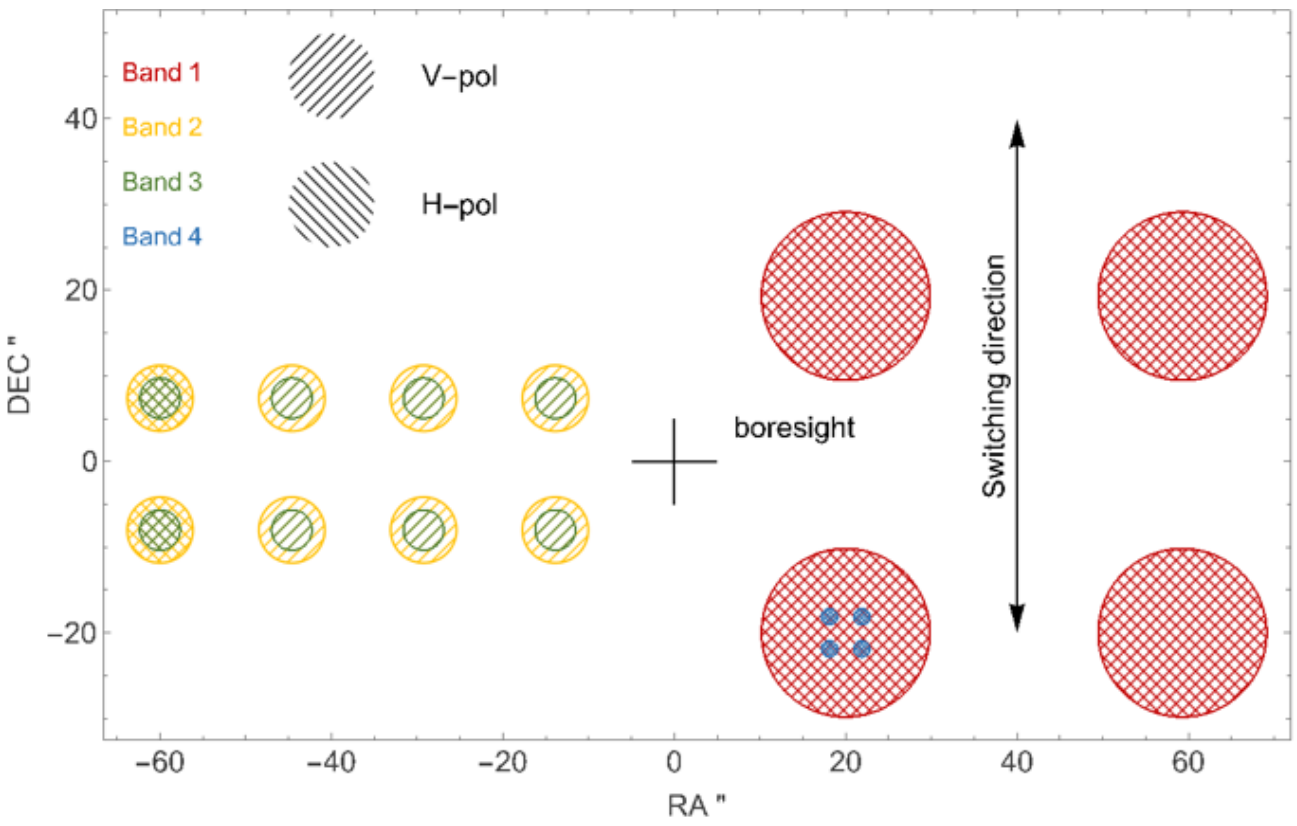

Figure 4: Layout of the beams from the four bands as imaged on the sky. The cross indicates the location of the telescope boresight. The beam sizes are calculated for the centre frequency of the four bands. Note that bands 1 and 4 have two polarizations for every pixel.

Bands 2 and 3 which are primarily used for large scale mapping will employ one array of 2x4 mixers and one orthogonally polarized 2x1 mixer array, i.e. a total of 10 mixers per band. This will enable the large-scale mapping with 8 on-sky beams, as well as point source observations chopping between two dual polarization beams. Should the on-board resources in the final payload design allow for more operating detectors the 2x1 mixer arrays will be extended to 2x4 arrays to provide increased sensitivity with 8 dual polarization beams per band allowing significantly faster mapping of the galactic plane.

| Band | B1 (SIS) | B2 (HEB) | B3 (HEB) | B4 (HEB - QCL) |
|---|---|---|---|---|
| Frequency Coverage (THz) | 0.45 – 0.65 | 1.11 -1.68 | 1.88 – 2.70 | 4.70 & 5.35 |
| Wavelength (µm) | 365 - 666 | 178 - 270 | 111 - 159 | 56 & 64 |
| DBS Receiver Sensitivity (K) | 50 – 100 | 370 - 480 | 480 - 680 | 1900 - 2200 |
| IF Bandwidth (GHz) | 6.5 | 4 | 4 | 3.5 |
| | | | | |
| Pixel Layout H/V Polarizations | H: 2x2 V: 2x2 | H: 2x4 V: 2x1 | H: 2x4 V: 2x1 | H: 2x2 V: 2x2 |
| Centre of Band Beamsize (") | 37 | 16 | 10 | 4.4 |
| Targeted Species | HDO, $H_2^{18}O$, $H_2O$, $NH_3$,$^{13}CO$, [CI]609 | [NII]205, $H_2O$ | [CII]158,HD, OH119, [NII]122 | [OI]63, HD |

Table 1: Specifications of the LETO configuration for the 4 receiver bands.

Bands 1 and 4 are mainly dedicated to point source observations and small maps and utilize two 2x2 arrays, thus enabling observations with 4 dual polarized beams. The specifications of the LETO bands together with the key astronomical lines they cover are listed in Table 1

### 3.4 Cold Focal Plane Unit

The cryogenic Focal Plane Unit houses the mixer arrays, local oscillator injection optics, calibration subsystem, and relay optics. To reduce mass and cryogenic volume, several options to combine bands into common units have been investigated. Since bands 2 and 3 will be used in parallel for the mapping of the galactic plane/external galaxies in the [CII] 158 µm (band 3) and [NII] 205 µm (band 2) lines, in the baseline configuration these two bands, will be combined in one optomechanical unit, including a common Offner relay and fore optics. Combining these bands in one unit will allow all the beams of the mixers of the two bands to be coaligned on the sky by use of a dichroic in the fore optics for the separation of the signal in the two frequency bands. In this design, incoming radiation from the telescope is relayed through the Offner optical system that incorporates a cryogenic chopper. The chopper selects between on-source observations, off-source sky positions, and internal hot and cold calibration loads. This design is derived directly from the successful Herschel/HIFI architecture and provides both short-term gain stability and absolute radiometric calibration.

Within each dual-band module, dichroic beam splitters separate the frequency bands while wire-grid polarizers divide the signal into orthogonal polarizations before coupling to the mixer arrays. The architecture enables precise co-alignment of the Band 2 and Band 3 beams, allowing simultaneous mapping of [C II] and [N II] emission across the Galactic plane.

### 3.5 Local Oscillator Subsystems

The local oscillator system is tailored to the requirements of each receiver band. For Bands 1–3, frequency generation is based on amplifier–multiplier chains employing Schottky-diode multiplier technology. These systems have extensive heritage from Herschel/HIFI, SOFIA/upGREAT, and JUICE/SWI. The multiplier chains provide broad frequency coverage, high reliability, and sufficient output power to drive multi-pixel arrays.

Band 4 operates above 4 THz where conventional multiplier chains are no longer practical. Therefore the local oscillator system employs quantum cascade lasers (QCLs) operating at 4.744 and 5.331 THz. These devices provide the only viable local oscillator technology at such frequencies and are derived from developments for GUSTO, SOFIA, and OSAS-B. Frequency stabilization is achieved through harmonic locking using Schottky devices, while beam multiplexing optics distribute the local oscillator power to the detector arrays.

### 3.6 Intermediate Frequency and Backend Electronics

The down-converted signals are amplified through a multi-stage intermediate-frequency chain. The first-stage amplifiers operate at 20 K to minimize system noise, while subsequent amplification stages are located at 120 K. The amplified signals are then transferred to the warm backend electronics housed within the service module.

Signal processing is performed by the Signal Processing Unit using HIFAS ASIC spectrometers developed by AAC Omnisys. Two complementary observing modes are provided. A wideband mode delivers 5 GHz instantaneous bandwidth with 5 MHz spectral resolution, while a high-resolution mode provides 0.2 MHz spectral resolution over a narrower spectral window. This dual-backend architecture allows LETO to support both large-scale survey observations and detailed kinematic investigations.

The Instrument Control Unit serves as the central command and monitoring system for the payload. Utilising dedicated subsystem controllers it manages receiver operation, local oscillator tuning, calibration sequences, thermal control, housekeeping monitoring, and science data packaging prior to transmission to the spacecraft mass memory.

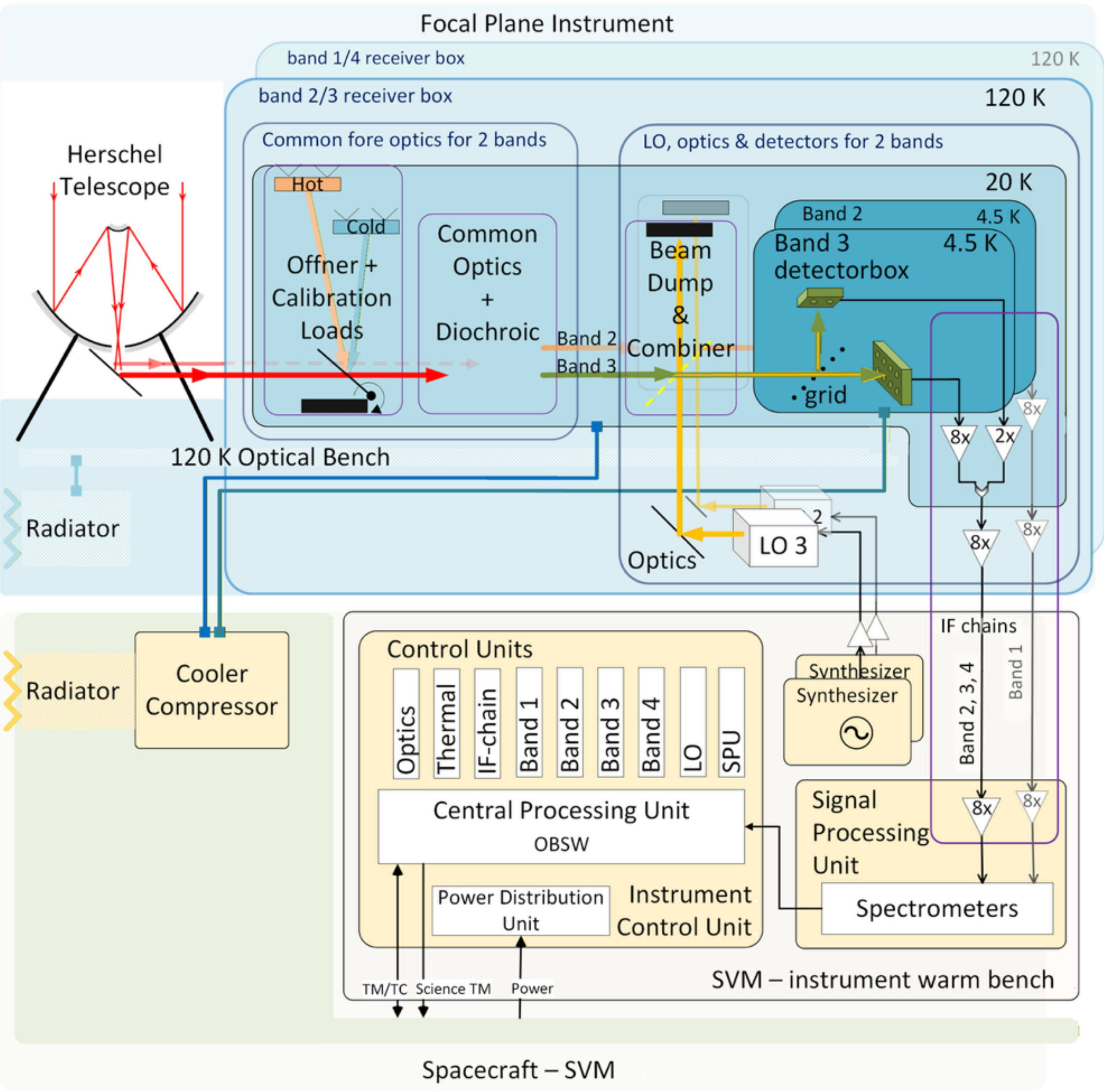

Figure 5: Functional block diagram of the LETO heterodyne spectrometer. Here the band 3 pixel configuration is shown – one polarization 2x8 mixers, orthogonal polarization 2x1 mixers. Band 2 has the same configuration, while bands 1 and 4 have 2x2 arrays for each polarization.

### 3.7 Thermal Architecture

The LETO thermal design combines passive and active cooling stages to support the various payload subsystems. The cryogenic architecture is centred on an ESA-provided multi-stage Cryo-Cooler Assembly delivering stable cooling at 4.5 K, 20 K, and 120 K. The 4.5 K stage supports the SIS and HEB mixer assemblies, while the 20 K stage accommodates the low-noise amplifiers. Intermediate optics and local oscillator components are located at 120 K (see also Figure 5). An additional dedicated 50 K cryocooler is included to cool the Band 4 quantum cascade laser subsystem.

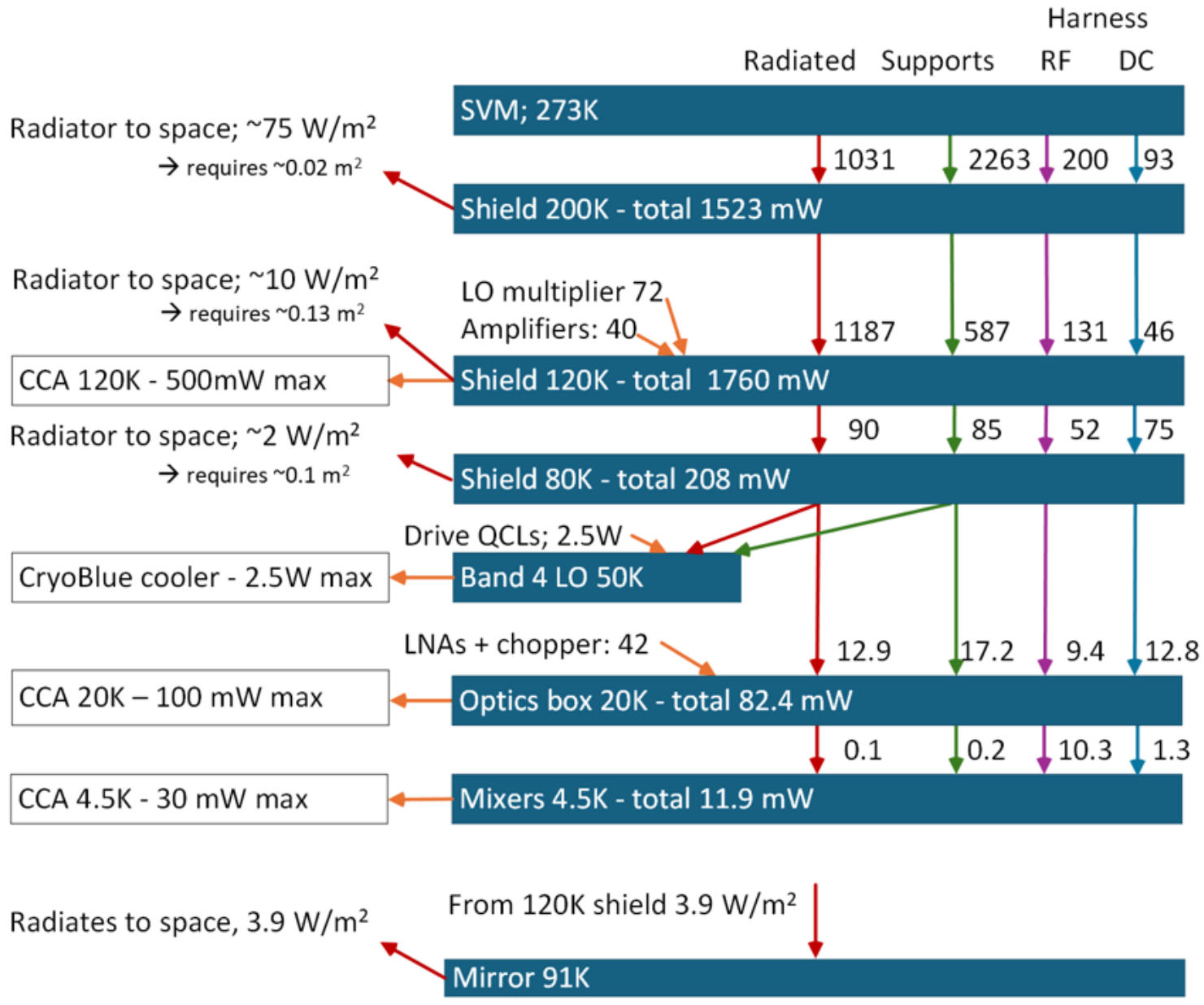

Figure 6: The LETO payload heat flow. Values in mW unless otherwise marked. Totals are the total heat that needs to be cooled by cooler or radiators, i.e. the "in" minus "out" from the radiated, support and harnesses.

The design minimizes parasitic heat loads and maintains thermal stability across the focal plane assembly through the use of low-emissivity shielding, low-conductance structural supports (e.g., CFRP), and carefully selected RF/DC harness materials. Thermal analyses, informed by heritage data from JWST/MIRI, demonstrate that conductive and radiative heat loads can be effectively managed. Additional thermal intercept stages at 80 K and 200 K, cooled by passive radiators, further reduce heat transfer to the colder cryogenic stages, while a small passive radiator supplements the 120 K cooling stage. The thermal interface between the ESA Cryo-Cooler Assembly and the focal plane will be optimized during Phase 0/A using high-conductivity thermal straps to minimize temperature gradients. Based on the Herschel telescope design and an estimated spacecraft heat load of ~3 $W/m^{-2}$, the LETO Telescope Assembly is expected to reach a stable operating temperature of approximately 100 K.

## 4. MISSION PROFILE AND LAUNCH SEGMENT

Operation at the Sun–Earth L2 Lagrange point provides an optimal environment for the LETO mission, offering both thermal stability and a nearly constant distance of approximately 1.5 million kilometres from Earth. From this location, the spacecraft can continuously point its telescope in directions perpendicular to the Sun-spacecraft vector, providing instantaneous access to a 360° annulus on the sky. As the spacecraft orbits the Sun, this viewing zone sweeps across the celestial sphere, enabling complete sky coverage within approximately six months. Maintaining the telescope boresight perpendicular to the Sun also ensures effective protection from direct solar illumination through the spacecraft sunshield while simultaneously providing a stable orientation for solar power generation.

Several orbit configurations around L2 are possible for LETO. The current baseline assumes a large-amplitude quasi-halo orbit similar to that used by Herschel, although the final orbit design will be refined during future mission phases. This option is favoured over a small-amplitude Lissajous orbit because it requires a lower acquisition $\Delta V$ and avoids eclipses, thereby simplifying thermal and power management. Additional factors, including communications performance, will be assessed during Phase 0 trade-off studies. The baseline launcher is Ariane 62, which can deliver approximately 3500 kg to an L2 transfer trajectory. If Ariane 64 becomes available, its greater launch capability could enable a dual-launch scenario similar to Herschel and Planck, potentially reducing mission costs. Based on the Herschel experience, LETO is expected to perform a series of trajectory-correction manoeuvres after separation from the launcher to acquire its operational orbit around L2, with the quasi-halo configuration requiring a relatively modest $\Delta V$ budget compared to alternative orbit options.